\documentclass[prl,aps,twocolumn,longbibliography]{revtex4-2}

\usepackage[paperwidth=210mm,paperheight=297mm,centering,hmargin=1.63cm,vmargin=2.2cm]{geometry}

\pdfoutput=1

\newcommand{\ac}{\alpha_{\textrm c}}

\newcommand{\tc}{t_{\textrm c}}
\newcommand{\tinf}{t_{\textrm s}}

\newcommand{\dNtc}{\Delta N_{\tc}}
\newcommand{\Ndotmax}{\dot{N}_{\rm max}}
\newcommand{\Ni}{N_0}
\newcommand{\lossL}{\mathcal{L}_3}

\newcommand{\ts}{t_{\textrm{s}}}

\newcommand{\nd}{n_{\textrm{d}}}

\newcommand{\rd}{r_{\textrm{d}}}

\newcommand{\td}{t_{\textrm{d}}}

\newcommand{\avdw}{r_{\rm vdW}}

\usepackage{float}
\usepackage{fourier}
\usepackage{amsmath,amssymb}
\usepackage{mathtools}
\usepackage{amsthm}
\usepackage{empheq}
\usepackage{cases}
\usepackage{times}
\usepackage{upgreek}
\usepackage{psfrag} 
\usepackage{latexsym} 
\usepackage{amstext}
\usepackage{amsfonts}
\usepackage{amsxtra} 
\usepackage{dcolumn}
\usepackage{dcolumn}
\usepackage{textcomp}
\usepackage{graphicx}
\usepackage{bm}
\usepackage{braket}
\usepackage{units}
\usepackage{lineno}
\usepackage[svgnames,dvipsnames]{xcolor}
\usepackage{booktabs}
\usepackage[normalem]{ulem}

\definecolor{myColor}{rgb}{0.02,0.12,0.3}
\definecolor{myciteColor}{rgb}{0.39,0.7,0.89}
\usepackage[colorlinks=true,citecolor=myColor,linkcolor=myColor,urlcolor=myColor]{hyperref}

\def\be{\begin{equation}}
\def\ee{\end{equation}}

\def\nobreakbefore{%
  \relax\ifvmode\else
    \ifhmode
      \ifdim\lastskip > 0pt\relax
        \unskip\nobreakspace
      \else 
        \nobreakspace
      \fi
    \fi
  \fi
}
\let\oldcite\cite
\renewcommand\cite{\nobreakbefore\oldcite}

\makeatletter
\def\@fnsymbol#1{\ensuremath{\ifcase#1\or *\or \dagger\or \ddagger\or
   \mathsection\or \mathparagraph\or \|\or **\or \dagger\dagger
   \or \ddagger\ddagger \else\@ctrerr\fi}}
\makeatother

\begin{document} 
 
\title{
Scaling Laws Governing the Collapse of a Bose--Einstein Condensate
}
\author{Sebastian~J.~Morris}
\author{Christopher J.~Ho}
\author{Simon M.~Fischer}
\author{Ji\v r\' i~Etrych}
\author{Gevorg~Martirosyan}
\author{Zoran~Hadzibabic}
\author{Christoph~Eigen}
\email{ce330@cam.ac.uk}
\affiliation{
Cavendish Laboratory, University of Cambridge, J. J. Thomson Avenue, Cambridge CB3 0HE, United Kingdom}

\begin{abstract}
We study the collapse of an attractive Bose--Einstein condensate, where an unstable system evolves towards a singularity, by numerically solving the underlying cubic-quintic nonlinear Schr\"odinger equation.
We find good agreement between our simulations and the atom-loss measurements with a $^{39}$K condensate.
Our simulations reveal an interplay of weak collapse and the propensity of the system to form a hotspot, and we uncover new scaling laws that govern this behavior.
We also identify promising signatures of the theoretically predicted, but so far experimentally elusive, elastic three-body interactions.
\end{abstract}
\maketitle

Collapse of nonlinear systems is ubiquitous in nature,
from the breaking of ocean waves~\cite{Davey:1974,McAllister:2024} to the gravitational collapse of stars~\cite{Penrose:1965}.
When attractive forces overwhelm dispersive ones, systems evolve towards a singularity in finite time.
In practice, dissipative processes typically become relevant as the singularity is approached, and can stabilize the system.

The same cubic nonlinear Schr\"odinger equation is at the heart of the theoretical description of a variety of collapse phenomena~\cite{Sulem:1999,Fibich:2015}, including self-focusing of light~\cite{Boyd:2009}, collapse of Langmuir waves in plasmas~\cite{Zakharov:1972,Kim:1974,Wong:1984}, and white-caps on choppy water~\cite{Davey:1974}.
A key role is played by the dimensionality: lower-dimensional systems tend to feature soliton solutions~\cite{Scott:1973}, including the celebrated Townes soliton in 2D~\cite{Moll:2003,Chen:2020,BakkaliHassani:2021}, whereas 3D systems predominantly exhibit collapse. For the 3D collapse, there is a rich variety of self-similar solutions that predict how a singularity is approached, including nonpeaked solutions~\cite{Fibich:2015} and the counterintuitive weak collapse~\cite{Zakharov:1975,Zakharov:1986,Zakharov:2012}, 
where a stronger attractive nonlinearity leads to less dissipation.
In comparison, little is understood about the dynamics after the collapse~\cite{Sulem:1999,Fibich:2015}, with possibilities including further collapse events and the formation of a hotspot~\cite{Zakharov:1981,Malkin:1988,Vlasov:1989,Zakharov:1989,Malkin:1990,Kosmatov:1991,Lemesurier:2000,Zakharov:2012}, where an emergent effective potential sustains a wave-action flux into a dissipation region in which the wave remains localized for a prolonged time.

Ultracold atoms provide powerful analog simulators of this nonlinear Schr\"odinger equation; the sign and the strength of the cubic nonlinearity can be tuned using Feshbach resonances\cite{Chin:2010}, and inelastic three-body recombination naturally acts as a quintic dissipation mechanism.
Collapse experiments have been performed with atomic Bose--Einstein condensates in harmonic~\cite{Gerton:2000,Roberts:2001,Donley:2001,Cornish:2006,Altin:2011,Compton:2012,Huang:2024} and box traps~\cite{Eigen:2016} (see also~\cite{otherexp}), and both the criterion and the time for the collapse to occur have been understood. 
Moreover, by isolating single-collapse events, the weak nature of the 3D collapse was observed~\cite{Eigen:2016}.
However, the scaling of the atom loss with the strengths of the cubic and quintic nonlinearities did not follow Zakharov's analytic prediction~\cite{Zakharov:1986}, and was never fully understood.

In this Letter, we perform numerical simulations of collapse in the cubic-quintic nonlinear Schr\"odinger equation.
Focusing on single-collapse events, the extracted total loss agrees with the measurements with a box-trapped $^{39}$K Bose--Einstein condensate.
The time-resolved loss dynamics feature an interplay of weak collapse and the propensity of the system to form a hotspot, and we uncover new scaling laws that govern this behavior.
We also study the effects of the theoretically predicted, but experimentally elusive, elastic three-body interactions~\cite{Wu:1959,Braaten:1999,Braaten:2006,Tan:2008c,Zwerger:2019,Mestrom:2020}.
At their predicted strength, these interactions modify the single-collapse loss only slightly, at a level that is not resolved in the experiments.
However, we find promising signatures in the time-resolved loss dynamics, including significantly longer dissipation times and suppression of the hotspot behavior.

In dimensionless form, the nonlinear Schr\"odinger equation with an elastic cubic and dissipative quintic term reads:
\begin{equation}
i \frac{\partial {\psi}}{\partial {t}}=-\nabla^2{\psi}+\alpha|{\psi}|^2{\psi}-i\eta |{\psi}|^4{\psi}\,,
\label{eq:ndegpe}
\end{equation}
where $\alpha$ and $\eta >0$ are real.
We initially ($t=0$) normalize the wavefunction $\psi(r,t)$ to unity inside a spherical box of unit diameter and enforce the boundary condition $\psi=0$ at the trap walls.
To convert from dimensionless units to the experimental ones for the collapse of a box-trapped condensate~\cite{Eigen:2016}, one rescales $r\to r/L$, $\psi \to L^{-3/2} \psi$, and $t\to t/\tau$, where $L$ is the characteristic box size, $\tau=2mL^2/\hbar$ the characteristic timescale, and $m$ the atom mass, and uses
\be
    \alpha=\frac{8\pi a \Ni}{L}\,,\ \ \eta=\frac{\Ni^2 m \lossL}{\hbar L^4}\,, 
    \label{eq:expunits}
\ee
where $a$ is the $s$-wave scattering length, $\Ni$ the initial condensate atom number, and $\lossL$ the three-body loss coefficient.

\begin{figure}[t!]
\centerline{\includegraphics[width=1\columnwidth]{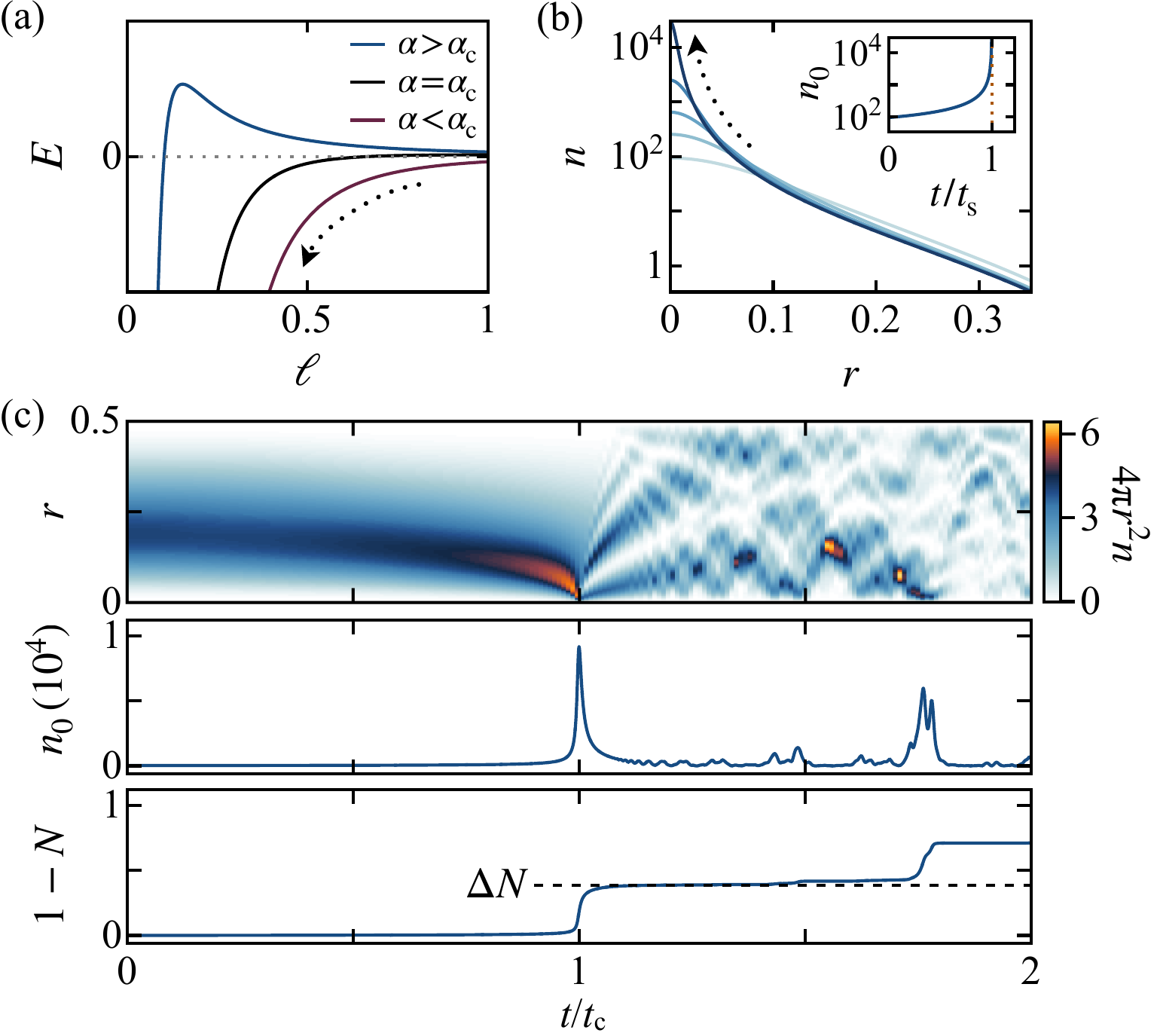}}
\caption{
Weak collapse.
(a)~Sketch of the effective energy landscape for a wavefunction of extent~$\ell$.
The interaction energy is $\propto \alpha/\ell^3 < 0$ and collapse occurs (arrow) for $\alpha<\ac$.
(b)~Typical evolution of the density profile $n(r,t)$ after a quench of $\alpha$ from above to below $\ac$.
As the system approaches the singularity (light to dark colors), the central density diverges at a time $\tinf$ (see inset), and a diminishing fraction of the wave approaches the singularity, with stationary tails left behind.
(c)~
Simulation results for a spherical box with $\ac=-3.89(1)$, $\alpha=-4.2$, and nonzero dissipation strength $\eta=3\times 10^{-4}$~[see Eq.~(\ref{eq:ndegpe})]. We show the dynamics of the radial population density $4\pi r^2 n$~(top), the central density $n_0$ (middle), and the fractional loss $1-N$~(bottom), where $N=\int 4\pi r^2 n\, {\rm d}r \leq 1$; $\tc$ is the time when $n_0$ first peaks (note that $\tc\to\ts$ for $\eta\to0$~\cite{tcvsts}).
The dashed line indicates $\Delta N$, the total single-collapse loss.
}
\label{fig1}
\vspace{-0.5em}
\end{figure}

Assuming $\eta=0$ and a characteristic wavefunction extent $\ell$ in Eq.~(\ref{eq:ndegpe}), the kinetic energy is $\propto 1/\ell^2$ and the interaction energy is $\propto \alpha/\ell^3$.
As sketched in Fig.~\ref{fig1}(a), for slightly negative $\alpha$ collapse is prevented by a kinetic energy barrier, which disappears at a negative critical value $\ac$.

As illustrated in Fig.~\ref{fig1}(b), after $\alpha$ is quenched (at $t=0$) from above to below $\ac$, the system approaches a singularity in time $\tinf$.
The density profile, $n(r,t)=|\psi(r,t)|^2$, is predicted~\cite{Zakharov:1986}, for $t \to \tinf$, to evolve self-similarly with a peaked scaling solution $\chi$: 
\be
n(r,t)=n_0(t) \chi[r/r_0(t)]\,,
\label{eq:scalingsoln}
\ee
with $\chi(0)=1$, $\chi \propto (r_0/r)^2$ for $r \gg r_0$, $r_0\propto \sqrt{\tinf - t}$, and $n_0(t) = n(0,t) \propto 1/[|\alpha|(\tinf - t)]$.
The collapse is termed \emph{weak} because the fraction of the wave that is collapsing towards the origin, $n_0 r_0^3 \propto \sqrt{\tinf - t}/|\alpha|\propto 1/\sqrt{n_0 |\alpha|^3}$, diminishes as $n_0$ diverges, with stationary tails, $n(r) \propto 1/(|\alpha| r^2)$, left behind~(see also~\cite{Supplemental}).

Introducing weak dissipation, $\eta \ll |\alpha|$, the quintic term in Eq.~(\ref{eq:ndegpe}) significantly affects the dynamics once it becomes comparable to the cubic one, which happens for $n_0 \propto |\alpha|/\eta$ and corresponding $r_0 \propto \eta^{1/2}/|\alpha|$.
At this point in time, the proportion of the wave that is still collapsing towards the origin is $\propto \eta^{1/2}/\alpha^2$. 
Assuming that this part is lost due to dissipation, while the tails are unaffected, gives the Zakharov scaling law for the fractional loss $\Delta N$ in a single-collapse event~\cite{Zakharov:1972}:
\begin{equation}
    \Delta N 
    \propto \eta^{1/2}/\alpha^2 \, ,
    \label{eq:weakcollapseloss}
\end{equation}
which, counterintuitively, is smaller for a more unstable system (larger $|\alpha|$).
These scaling arguments provide some intuition, but a complete understanding of collapse in Eq.~(\ref{eq:ndegpe}) is lacking\cite{Fibich:2015}, and it is in general difficult to simulate due to the necessary spatiotemporal resolution~\cite{LeMesurier:1988,Kosmatov:1991,Bao:2013}.

Here, we assume spherical symmetry~\cite{sphericalsymmetry} and numerically solve the radial part of Eq.~(\ref{eq:ndegpe}) in 1D~(see \cite{Supplemental} for details).
We mimic the experimental protocol from Ref.~\cite{Eigen:2016} by starting from the metastable state for $\alpha = -3.6$, above $\ac\approx -3.9$. We then quench to $\alpha < \ac$, and record the evolution of~$\psi$ for different values of $\eta$.

In Fig.~\ref{fig1}(c) we show a typical example of the simulated dynamics, which reveals an isolated first collapse event, followed by rich dynamics, including refocusing and subsequent collapses.
We plot the evolution of the radial population density $4\pi r^2 n(r,t)$ (top), the central density $n_0$ (middle), and the loss $1-N(t)$ (bottom), where $N(t)=\int 4\pi r^2 n(r,t)\,{\rm d}r \leq 1$.
Here $\tc$ is the time when $n_0$ first peaks~\cite{tcvsts} and the dashed line shows the extracted $\Delta N$; note that throughout this Letter we focus on the first collapse event.

\begin{figure}[t!]
\centerline{\includegraphics[width=1\columnwidth]{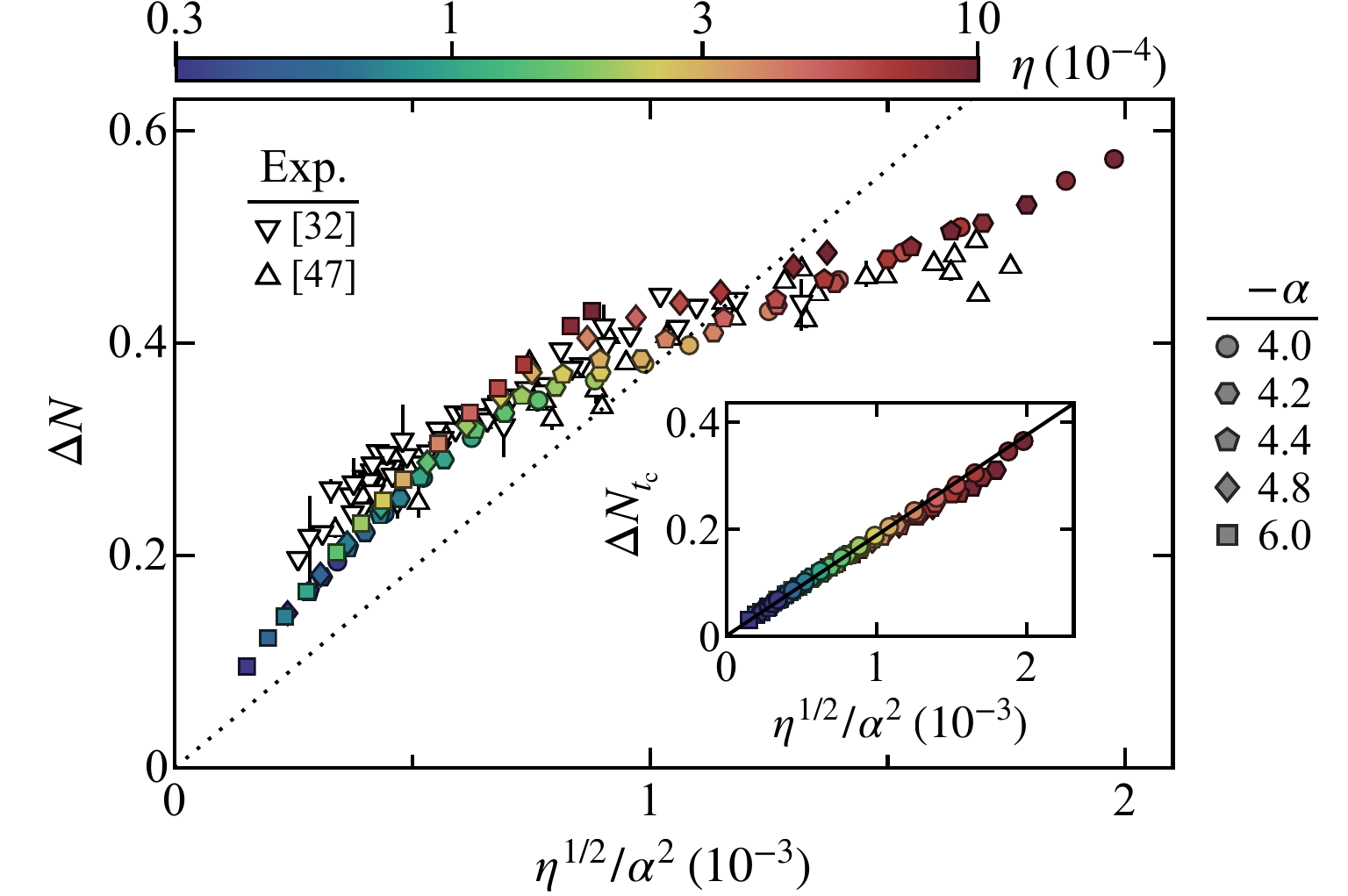}}
\caption{
Single-collapse atom-loss scaling laws.
Plot of $\Delta N$ versus $\eta^{1/2}/\alpha^2$, comparing the experimental data (open triangles) to our simulations with different $\alpha$ (colored symbols) and $\eta$ (color bar).
The inset shows the simulated $\dNtc$, the loss up to $\tc$, which follows the prediction in Eq.~(\ref{eq:weakcollapseloss}) (solid line), and suggests that the different scaling of $\Delta N$ arises due to dynamics in the aftermath of the singularity; the dotted line shows $\Delta N = 2\dNtc$.
Note that the systematic uncertainty in experimentally determining $\eta^{1/2}/\alpha^2$ is $\pm20\%$.
}
\label{fig2}
\vspace{-1em}
\end{figure}

\begin{figure*}[t!]
\centerline{\includegraphics[width=1\textwidth]{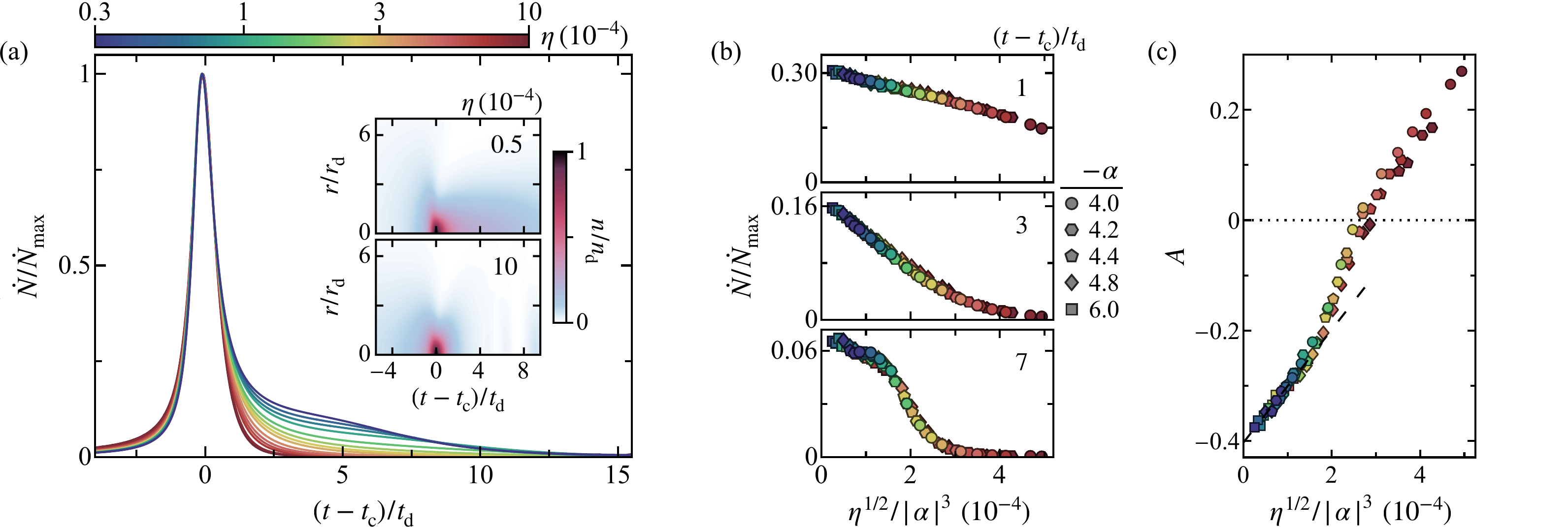}}
\caption{
Time-resolved loss dynamics revealing the scaling laws that govern the propensity to form a hotspot.
(a)~Normalized loss rate $\dot{N}/\Ndotmax$ versus $(t-\tc)/\td$ for $\alpha=-4.0$ and varying $\eta$; here $\td$ is the full-width-half-maximum of $|\dot{N}(t)|$. 
For smaller $\eta$, the universal weak-collapse peak is accompanied by a growing `shoulder' (for $t > \tc$).
The insets show $n(r,t)$ for $\eta= 5\times 10^{-5}$ (top) and $\eta= 10^{-3}$ (bottom), where for the smaller $\eta$ the density remains high and localized for a prolonged time, a key signature of hotspot behavior; here $\nd=n(0, \tc)$ and $\rd$ is defined so that $n(\rd, \tc) = \nd/2$. 
(b) Plotting $\dot{N}/\Ndotmax$ at three characteristic $(t-\tc)/\td>0$ (panels) versus $\eta^{1/2}/|\alpha|^3$ for different $\alpha$ (legend) and $\eta$ [colors, as in (a)] reveals remarkably universal behavior.
(c) The loss asymmetry $A=2\dNtc/\Delta N-1$ also shows universal behavior in $\eta^{1/2}/|\alpha|^3$. 
For $\eta^{1/2}/|\alpha|^3\to 0$ we find that $A$ approaches $\approx-0.4$; the dashed line shows a linear fit for $\eta^{1/2}/|\alpha|^3<1\times 10^{-4}$.
}
\label{fig3}
\vspace{-1em}
\end{figure*}

In Fig.~\ref{fig2}, we compare our results for $\alpha\in [-4,-6]$ and $\eta\in[0.3, 10] \times 10^{-4}$ (colored symbols) to the measurements of $\Delta N$ with a box-trapped $^{39}$K condensate~\cite{Eigen:2016,Supplemental} (open triangles), finding good agreement.
In both simulation and experiment, the data do not follow the prediction $\Delta N\propto \eta^{1/2}/\alpha^2$~[Eq.~(\ref{eq:weakcollapseloss})].
However, as shown in the inset, the loss up to $\tc$, denoted $\dNtc$, does show this scaling (solid line), so the difference must occur after $\tc$ (cf.~dotted line, which shows $\Delta N = 2\dNtc$).

To elucidate the collapse dynamics, we look at the time-resolved loss rate $\dot{N}={\rm d}N/{\rm d}t=-8\pi \eta \int n^3 r^2{\rm d}r$.
In Fig.~\ref{fig3}(a), for $\alpha=-4.0$ and various $\eta$, we plot $\dot{N}/\Ndotmax$ versus $(t-\tc)/\td$, where $\Ndotmax$ is the maximal instantaneous loss rate and $\td$ the dissipation time, defined as the full-width-half-maximum of $|\dot{N}(t)|$.

Before and near $\tc$, the curves coincide, meaning that $\dNtc\propto \td |\Ndotmax|$. This is consistent with the Zakharov weak-collapse picture~\cite{Kosmatov:1991}, with the characteristic size of the dissipation region $\rd\propto r_0(\tc)\propto\eta^{1/2}/|\alpha|$, density $\nd=n_0(\tc)\propto|\alpha|/\eta$, and time $\td\propto\eta/\alpha^2$, such that $\td |\Ndotmax|\propto\td\eta\nd^3\rd^3\propto\eta^{1/2}/\alpha^2$, as in Eq.~(\ref{eq:weakcollapseloss}).
We define $\rd$ so that $n(\rd,\tc)=\nd/2$, and have verified the scaling laws for $\nd$, $\rd$, $\td$, and $\Ndotmax$~(see~\cite{Supplemental}).

However, for the smaller $\eta$ values the loss extends to longer times, with a prominent `shoulder' at $t>\tc$.
As illustrated in the inset, the corresponding density profiles remain localized for much longer ($\gtrsim 5\td$), which is a key signature of hotspot formation~\cite{Zakharov:1981,Malkin:1988,Vlasov:1989,Zakharov:1989,Malkin:1990,Kosmatov:1991,Lemesurier:2000,Zakharov:2012}~(see also~\cite{prolongedburningdistinction}).

In Fig.~\ref{fig3}(b) we extend our analysis to different $\alpha$ values and show that the hotspot shoulder exhibits remarkably universal behavior: at a fixed \mbox{$(t - \tc)/\td>0$}, we empirically find that $\dot{N}/\Ndotmax$ is a function of $\eta^{1/2}/|\alpha|^3$~\cite{deltanuniversality}.
To our knowledge, this scaling has not been predicted, and invites further study~\cite{pasthotspots}.

In Fig.~\ref{fig3}(c) we introduce the loss asymmetry $A\equiv2\dNtc/\Delta N-1$, which is also a universal function of $\eta^{1/2}/|\alpha|^3$.
Interestingly, in the regime where the hotspot behavior is most pronounced ($\eta^{1/2}/|\alpha|^3\to~0$), $A$ approaches $\approx -0.4$ (rather than $-1$), so that $\Delta N \approx 3.3\, \dNtc$.
Note that this recovers the Zakharov scaling $\Delta N \propto \eta^{1/2}/\alpha^2$, and since $A$ grows linearly for small $\eta^{1/2}/|\alpha|^3$ (dashed line), the next-order correction to $\Delta N$ is $\propto\eta/|\alpha|^5$.
For large $\eta^{1/2}/|\alpha|^3$, $A$ exceeds~$0$ because a significant fraction of the wave is lost already before~$\tc$.

In experimental units, our simulations predict a size of the dissipation region $\rd L \approx 1.9\times (m \lossL/\hbar)^{1/2}/(8\pi |a|)$,  a peak density $\nd \Ni / L^3 \approx 0.72 \times 8\pi \hbar |a| / (m \lossL)$, and a dissipation time $\td \tau \approx 25\times m^2 \lossL / (32 \pi^2 \hbar^2 a^2)$~\cite{Supplemental}; note that these do not depend on either $\Ni$ or $L$.
For~\textsuperscript{39}K with $\lossL=1.3(5) \times 10^{-41}\textrm{m}^6\textrm{s}^{-1}$~\cite{Fattori:2008a} at $a=-0.3 a_0$~\cite{potassiumexperiment} (where $a_0$ is the Bohr radius), $\rd L\approx 0.4\,\upmu$m, $\nd \Ni / L^3 \approx4\times 10^{4}\,\upmu \text{m}^{-3}$, and $\td \tau\approx 2\,$ms.
By leveraging nondestructive imaging techniques~\cite{Andrews:1996,Meppelink:2010,Gajdacz:2013} to monitor the same cloud as it collapses, or by preparing samples with a small $\Ni$ uncertainty
~\cite{Gajdacz:2016} (to minimize shot-to-shot variations in $\tc$) it should be possible to experimentally resolve the collapse dynamics.

\begin{figure}[b!]
\centerline{\includegraphics[width=\columnwidth]{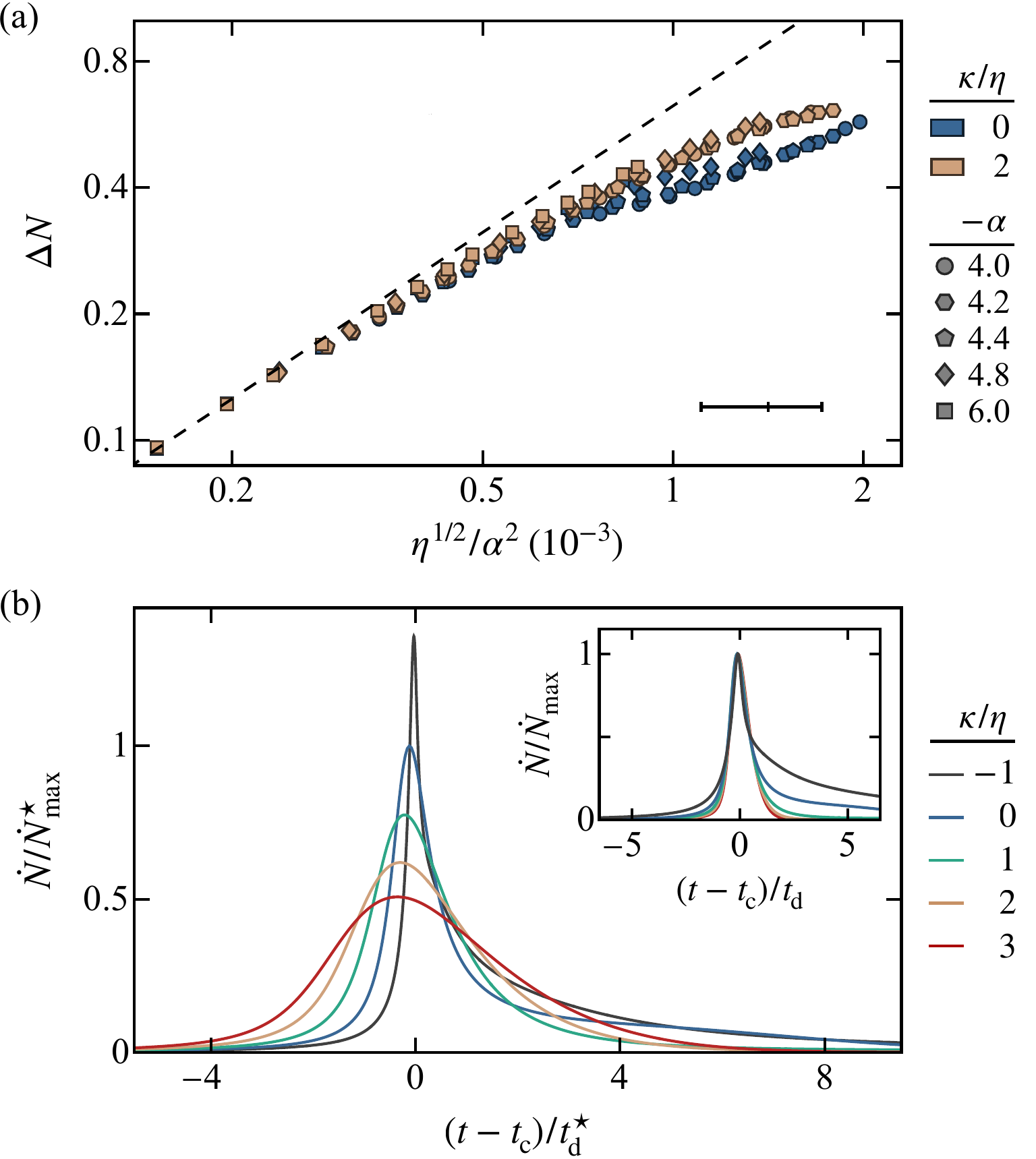}}
\caption{
Effects of elastic three-body interactions.
(a)~Plot of $\Delta N$ versus $\eta^{1/2}/\alpha^2$ for $\kappa/\eta=2$ and $\kappa=0$ (colors) on log-log scale, also varying both $\alpha$ (symbols) and $\eta\in[0.3, 10] \times 10^{-4}$ as in Fig.~\ref{fig2}.
For reference, the dashed line shows $\propto \eta^{1/2}/\alpha^2$, and the $x$ error bar in the bottom right corner indicates the $\pm 20\%$ systematic uncertainty in experimentally determining $\eta^{1/2}/\alpha^2$~\cite{alphackappashift}.
(b)~Time-resolved $\dot{N}$ for different $\kappa/\eta$ (legend) and fixed $\alpha=-4.4$ and $\eta=10^{-4}$, highlighting that the dissipation time $\td$ and hotspot behavior are highly sensitive to $\kappa/\eta$, even though the differences in $\Delta N$ are small.
The main panel shows $\dot{N}/\dot{N}^{\star}_{\rm max}$ versus $(t-\tc)/\td^{\star}$, where the stars ($^{\star}$) indicate $\kappa=0$ values, while the inset shows individually normalized curves.
}
\label{fig4}
\vspace{-1em}
\end{figure}

Note that beyond-mean-field effects~\cite{Lee:1957a}, which are not captured by Eq.~(\ref{eq:ndegpe}), are expected to be negligible in the current $^{39}$K experiments, since $\sqrt{(\nd \Ni/L^3) |a|^3} \approx10^{-4} (a/a_0)^2 \lesssim 3 \times 10^{-4}$.
However, by increasing $|a|$ while keeping $\alpha\approx\ac$, it should be possible to also probe these effects; e.g.~using $a=-30a_0$ would offer a sizable $\sqrt{(\nd \Ni/L^3) |a|^3} \approx 0.1$ at $\tc$. 

In the final part of this Letter, we extend our simulations to include elastic three-body interactions~\cite{Wu:1959,Braaten:1999,Braaten:2006,Tan:2008c,Zwerger:2019,Mestrom:2020}, predicted to accompany the inelastic ones in atomic gases~\cite{Zwerger:2019,Mestrom:2020}, by adding a real quintic term $\kappa|\psi|^4\psi$ to Eq.~(\ref{eq:ndegpe}), i.e.~$-i\eta\to -i\eta+\kappa$.
The strengths of the elastic and inelastic three-body interactions are, respectively, $\kappa = \text{Re}[D] N_0^2/L^4$ and $\eta = -\text{Im}[D] N_0^2/L^4$, where $D$ is the complex three-body scattering hypervolume~\cite{Tan:2008c}.
Near the zero crossing of a Feshbach resonance, Im$[D]=-\lossL m/\hbar$ is nonuniversal (as it depends on details of the short-range interactions) but it can readily be measured, whereas the elastic three-body interactions have been predicted to be universally set by the van-der-Waals interaction length $\avdw$, with $\text{Re}[D]\approx 100\avdw^4$~\cite{Zwerger:2019,Mestrom:2020}.
Combining this prediction with the independently measured $-\text{Im}[D]= 60(20)\avdw^4$ for $^{39}$K~\cite{Fattori:2008a} suggests $\kappa/\eta\approx 2$.

In Fig.~\ref{fig4}(a) we show $\Delta N$ versus $\eta^{1/2}/\alpha^2$ for both $\kappa / \eta = 2$ and $\kappa=0$ on log-log scale, with the same $\alpha$ and $\eta$ values as in Fig.~\ref{fig2}.
The differences in $\Delta N$ between the two are relatively small ($\lesssim 20\%$), and moreover, cannot be resolved within the systematic uncertainty of $\pm 20\%$ in experimentally determining $\eta^{1/2}/\alpha^2$.

However, as illustrated in Fig.~\ref{fig4}(b) for $\eta^{1/2}/\alpha^2 \approx 5\times 10^{-4}$ ($\alpha=-4.4$ and $\eta=10^{-4}$), 
the time-resolved loss dynamics are very sensitive to $\kappa$.
For increasing $\kappa$, the time $\td$ increases, $\Ndotmax$ decreases, and the hotspot behavior is suppressed.
Note that, when $\kappa > 0$, the elastic three-body term plays a similar role to the dissipative one in slowing down the approach to the singularity (see \cite{Supplemental} for additional details).

Experimental studies of the collapse dynamics in $^{39}$K could thus offer a setting for observing elastic three-body interactions.
To study the hotspot behavior, one could instead use other atoms with access to smaller predicted $\kappa/\eta$, e.g. for $^{7}$Li with $-\text{Im}[D]\approx 5000 \avdw^4$~\cite{Shotan:2014} one has $\kappa/\eta \approx 0.02$.

In conclusion, we numerically investigated collapse in the 3D cubic-quintic nonlinear Schr\"{o}dinger equation, finding good agreement between our simulations and experiments performed with attractive box-trapped Bose--Einstein condensates.
We demonstrated the propensity of the system to form a hotspot, uncovering new scaling laws that extend the Zakharov picture of weak collapse, and showed that collapse dynamics can be a sensitive probe of elastic three-body interactions.
In the future, it would be interesting to further study the interplay between elastic and inelastic three-body interactions, for instance in the formation of quantum droplets~\cite{Bulgac:2002,Zwerger:2019,Mestrom:2020,Son:2022,Hofmann:2022}.
It would also be interesting to study how the presence of a harmonic trap modifies the collapse dynamics, where soliton formation can play a role~\cite{Cornish:2006} and delayed collapses have been predicted~\cite{Biasi:2017}.
More generally, studying collapse dynamics of other quantum fluids, including quantum mixtures~\cite{Baroni:2024} and dipolar condensates~\cite{Chomaz:2022,Bigagli:2024}, could offer more insights into the interactions governing their behavior (see e.g.~\cite{Petrov:2014,Petrov:2015,Blakie:2016,Zwerger:2024}).

\section*{Acknowledgements}

We thank Wilhelm Zwerger, Robert P. Smith, Martin Gazo, and Servaas Kokkelmans for discussions and comments on the manuscript.
This work was supported by EPSRC [Grant No.~EP/P009565/1], ERC [UniFlat], and STFC [Grant No.~ST/T006056/1]. Z.H. acknowledges support from the Royal Society Wolfson Fellowship.

%


\setcounter{figure}{0} 
\setcounter{equation}{0} 

\renewcommand\theequation{S\arabic{equation}} 
\renewcommand\thefigure{S\arabic{figure}} 

\newpage
~
\newpage

\section*{Supplemental Material}


\begin{figure}[b!]
\centerline{\includegraphics[width=\columnwidth]{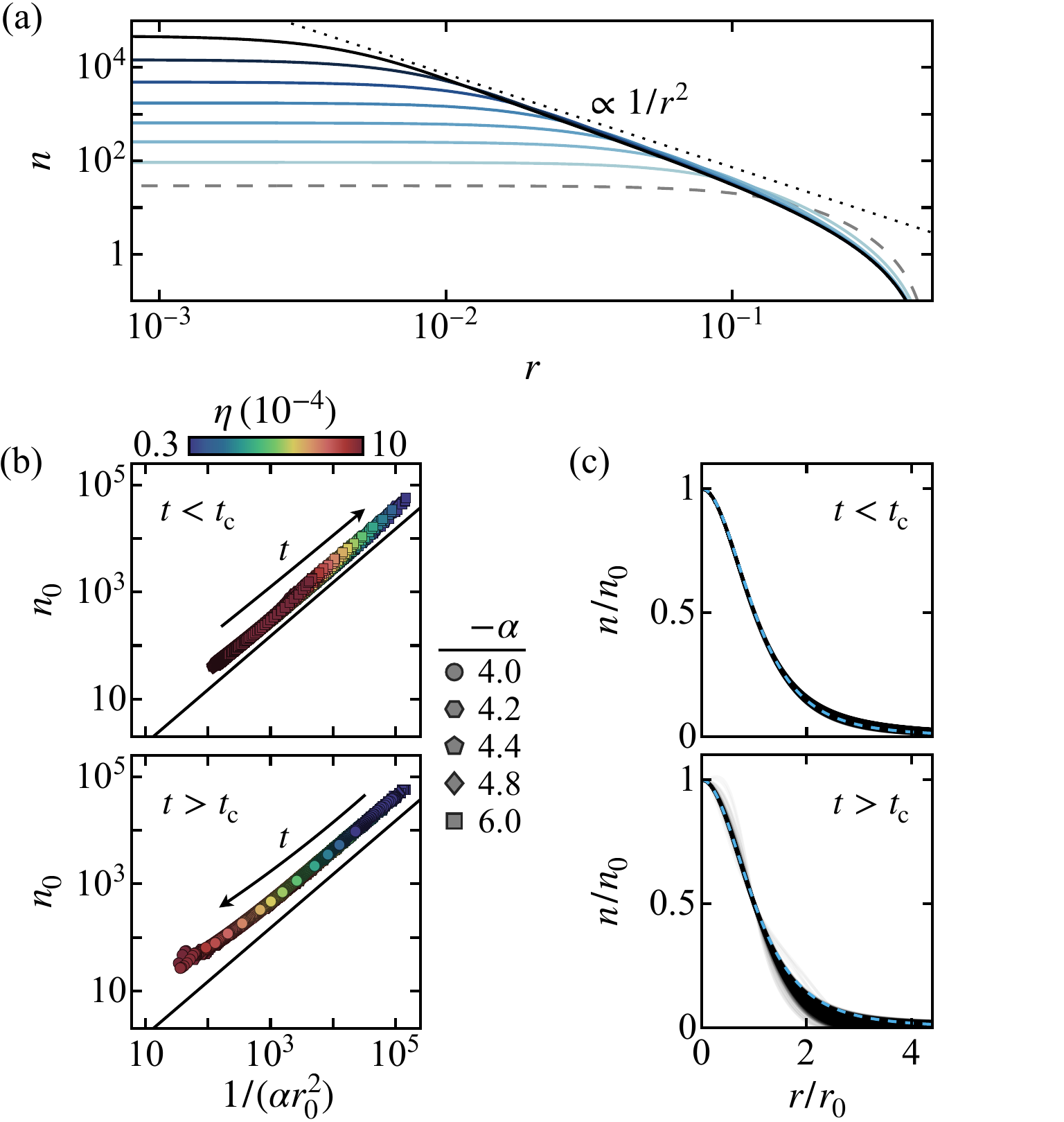}}
\caption{
Self-similar wavefunction evolution throughout the collapse.
(a)~Log-log plot of $n(r)$ at different times $t<\tc$ (light to dark colors), for $\alpha=-4$ and $\eta\approx0$.
The gray dashed line shows $n(r,0)$.
(b)~Plot of $n_0$ versus $1/(\alpha r_0)^2$ for all our $\alpha$ (legend) and $\eta$ (color bar), evenly sampled from $-5<(t-\tc)/\td<0$ (top) and $0<(t-\tc)/\td<5$ (bottom), revealing a universal scaling trajectory.
For reference, the solid lines show $n_0 \propto 1/(\alpha r_0^2)$. 
(c)~Plot of all the normalized density profiles $n(r/r_0,t)/n_0(t)$ for the same \mbox{$\alpha$, $\eta$, and $t$} as shown in (b). Each individual curve is shown with $3\%$ opacity and the dashed lines show squared Lorentzians.
}
\label{figS1}
\vspace{-1em}
\end{figure}

\subsection{\textsc{I. The Collapsing Wavefunction}}
\vspace{-1em}

In Fig.~\ref{figS1}(a), we show the evolution of the density profile $n(r,t)$ on log-log scale following a collapse-inducing quench to $\alpha=-4$ for $\eta\approx 0$, illustrating the weak-collapse picture.

In Fig.~\ref{figS1}(b), we summarize the dynamics of $n(r)$ near $\tc$, for both $-5 < (t - \tc)/\td < 0$ (top) and $0 < (t - \tc)/\td < 5$ (bottom), and our set of $\alpha\in [-4,-6]$ and $\eta\in[0.3, 10] \times 10^{-4}$ values.
Plotting $n_0$ versus $1/(\alpha r_0^2)$, where $r_0(t)$ is defined so that $n(r_0, t) = n_0/2$, shows that the relation between $n_0$ and $r_0$ is universal both before and after $\tc$.

In Fig.~\ref{figS1}(c), for the same times as shown in Fig.~\ref{figS1}(b), we plot all the normalized density profiles $n(r/r_0,t)/n_0(t)$, showing that for $r \lesssim 2r_0$ they are well captured by the same shape.
Note that here $\gtrsim 95\%$ of the instantaneous loss takes place at $r<2r_0$.

\subsection{\textsc{II. Numerical Simulations}}
We solve Eq.~(\ref{eq:ndegpe}) on a GeForce RTX 4070 GPU by assuming spherical symmetry and applying the time-splitting finite-difference method~\cite{Adhikari:2002,Bao:2013}, where we make use of the algorithm from Ref.~\cite{Valero:2018} to solve the relevant tri-diagonal matrix systems.

We discretize space inside the unit-diameter box into $2^{12}$ evenly spaced radii from $r=10^{-14}$ (to avoid divide-by-zero errors) to $r=0.5$, with the boundary condition $\psi(0.5, t) = 0$ applied during each timestep $\Delta t \approx 1.5 \times 10^{-8}$. 
We have verified that varying both the spatial and/or temporal sampling by a factor $2$ does not meaningfully affect our results.

\subsection{\textsc{III. New Experimental Data}}

\begin{figure}[b!]
\centerline{\includegraphics[width=\columnwidth]{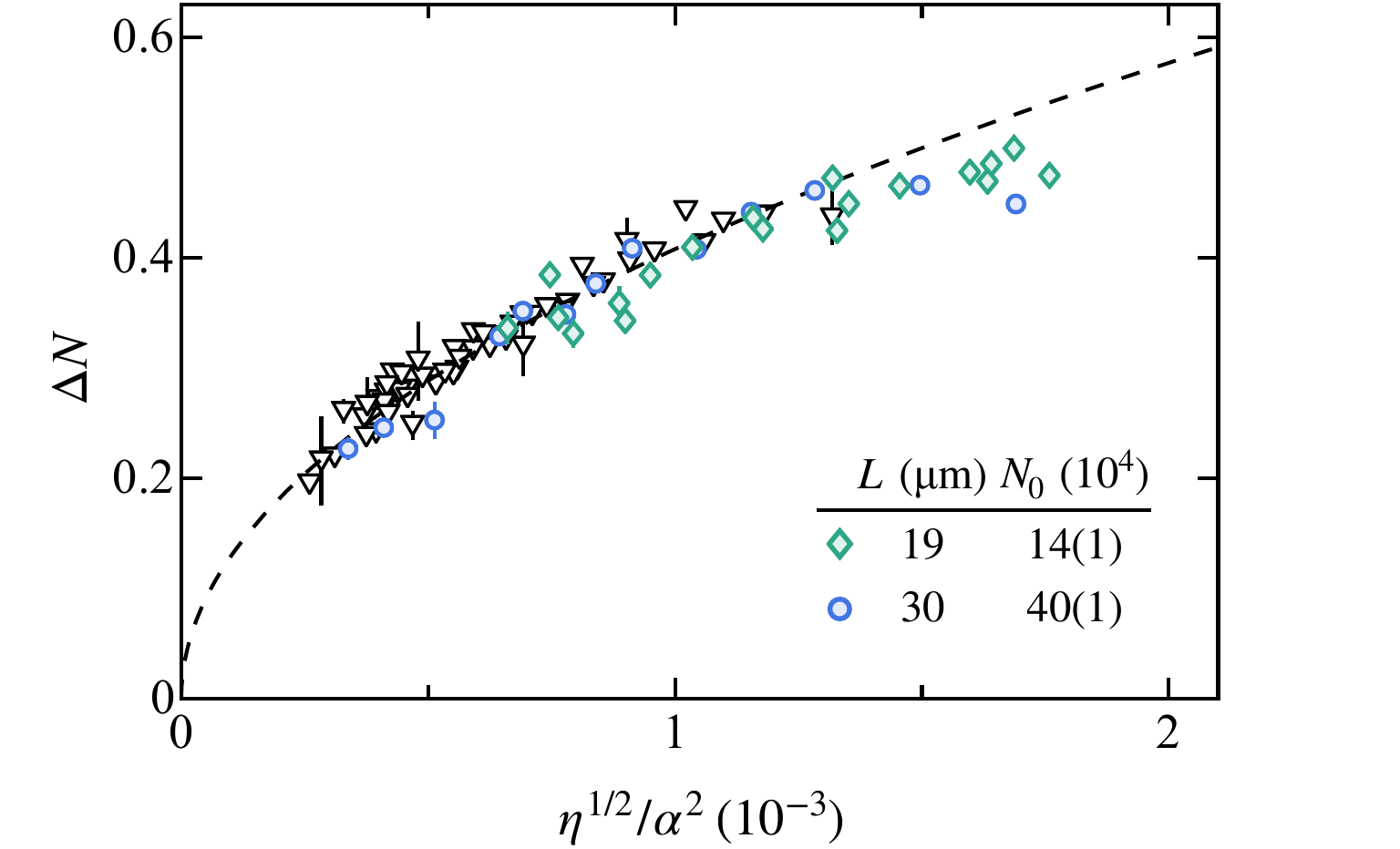}}
\caption{
New single-collapse loss measurements with $^{39}$K.
Plot of $\Delta N$ versus $\eta^{1/2}/\alpha^2$, showing the previous measurements from Ref.~\cite{Eigen:2016} (black triangles) alongside our new measurements (legend) that extend to larger values of $\eta^{1/2}/\alpha^2$.
For reference, the dashed line shows $\Delta N\approx 13\times \eta^{1/4}/|\alpha|$~\cite{Eigen:2016}.
}
\label{figS2}
\end{figure}

We perform additional measurements of $\Delta N$ as in Ref.~\cite{Eigen:2016}, extending the range of $\eta^{1/2}/\alpha^2$ to larger values.
We start with a quasipure $^{39}$K Bose--Einstein condensate consisting of $\Ni$ atoms in the lowest hyperfine state, confined in a cylindrical optical-box trap of radius $R$ and length $L\approx2R$, and tune the magnetic field to near the zero crossing at $350.4(1)$\,G~\cite{Fattori:2008a}.
Note that we always use the $1/\Ni$-dependence of the critical interaction strength $a_{\rm c}$ to define the effective $a=0$~\cite{Roberts:2001,Eigen:2016}.
To initiate collapse, we quench $a$ from above to below $a_{\rm c}$, and measure $\Delta N$.

In Fig.~\ref{figS2} we plot $\Delta N$ versus $\eta^{1/2}/\alpha^2$ (cf.~Fig.~\ref{fig2}), delineating the new data (blue circles and green diamonds) from previous measurements~\cite{Eigen:2016} (black triangles).
Note that for the blue circles we quench from $\alpha\approx0$ (instead of starting close to $\ac$~\cite{Eigen:2016}), demonstrating that $\Delta N$ is robust to the initial condition.
The $\Delta N$ measurements span $\alpha\in [-4,-11]$ and $\eta\in [0.4, 15]\times 10^{-3}$.
The key limitation for measuring $\Delta N$ for smaller $\eta^{1/2}/\alpha^2$ is the onset of multiple-collapse behavior~\cite{Eigen:2016},
while reaching larger $\eta^{1/2}/\alpha^2$ is complicated by three-body loss becoming relevant already for $t\ll\tc$ and also during sample preparation.

\vspace{-1em}
\subsection{\textsc{IV. Zakharov Scaling Laws}}
{
In Fig.~\ref{figS3}, we separately test the predictions for the different Zakharov weak-collapse scaling laws for $\nd$, $\rd$, $\td$, and $\Ndotmax$.
The solid lines show linear fits to data with small dissipation ($\dNtc < 0.2$ corresponding to $\eta^{1/2}/\alpha^2< 1.06\times 10^{-4}$), yielding:
$\nd \approx 0.72 \times |\alpha|/\eta$, $\rd \approx 1.9 \times \eta^{1/2}/|\alpha|$, $\td \approx 25\times \eta/\alpha^2$, and $|\Ndotmax| \approx 4.1\times \eta \rd^3 \nd^3 \approx 11 /\eta^{1/2}$.

In Fig.~\ref{figS4}, we provide additional details on the effects of elastic three-body interactions, by comparing $\kappa=0$ and \mbox{$\kappa/\eta=2$} for the same $\alpha$ and $\eta$ values as in Fig.~\ref{fig4}(a).

In Fig.~\ref{figS4}(a), we show that the Zakharov scaling laws for $\nd$, $\rd$, $\td$, and $\Ndotmax$ remain essentially universal for $\kappa/\eta=2$, while the proportionality constants differ significantly.
Qualitatively, one may expect the collapse to be arrested when $n_0=\nd$ such that $|\alpha| \nd \sim |\kappa -i\eta|\nd^2$, so introducing $\kappa > 0$ decreases $\nd$, $|\Ndotmax|$, $1/\rd$, and $1/\td$.
Note that for $\eta=0$ and $\alpha < 0$, exotic quantum droplets are predicted, where the repulsive three-body interactions can stabilize the system so long as $N$ is above a critical value $N_{\rm c}\propto \kappa^{1/2}/\alpha^2$~\cite{Zwerger:2019,Mestrom:2020,Son:2022}, which may suggest a continuous connection with the scaling $\Delta N \propto \eta^{1/2}/\alpha^2$ for $\kappa = 0$.
However, exploring the full dynamical phase diagram deserves a separate dedicated study~(see also~\cite{Son:2022,Hofmann:2022}).

As shown in Fig.~\ref{figS4}(b), for $\eta^{1/2}/|\alpha|^3 \to 0$, $|A|$ is significantly smaller for $\kappa/\eta=2$, which reflects the suppression of the hotspot behavior.
In Fig.~\ref{fig4}(c), we plot the normalized density profiles for $\kappa=0$ and $\kappa/\eta = 2$ at three different $(t - \tc)/\td$, which show differences for $t\gtrsim \tc$.

\onecolumngrid
~
\begin{figure*}[t!]
\centerline{\includegraphics[width=\columnwidth]{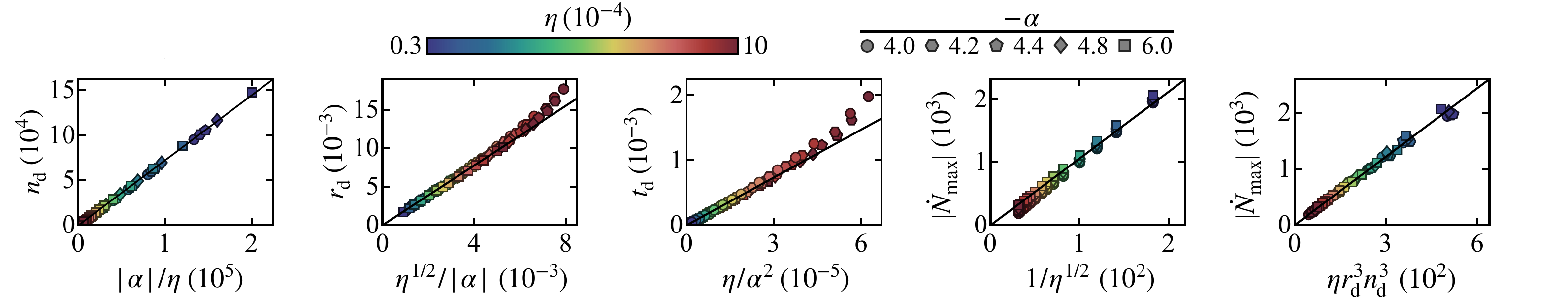}}
\caption{
Testing the Zakharov scaling law predictions (solid lines) for key quantities in our simulations, varying $\alpha$ and $\eta$.
}
\label{figS3}
\end{figure*}

\begin{figure*}[t!]
\centerline{\includegraphics[width=\columnwidth]{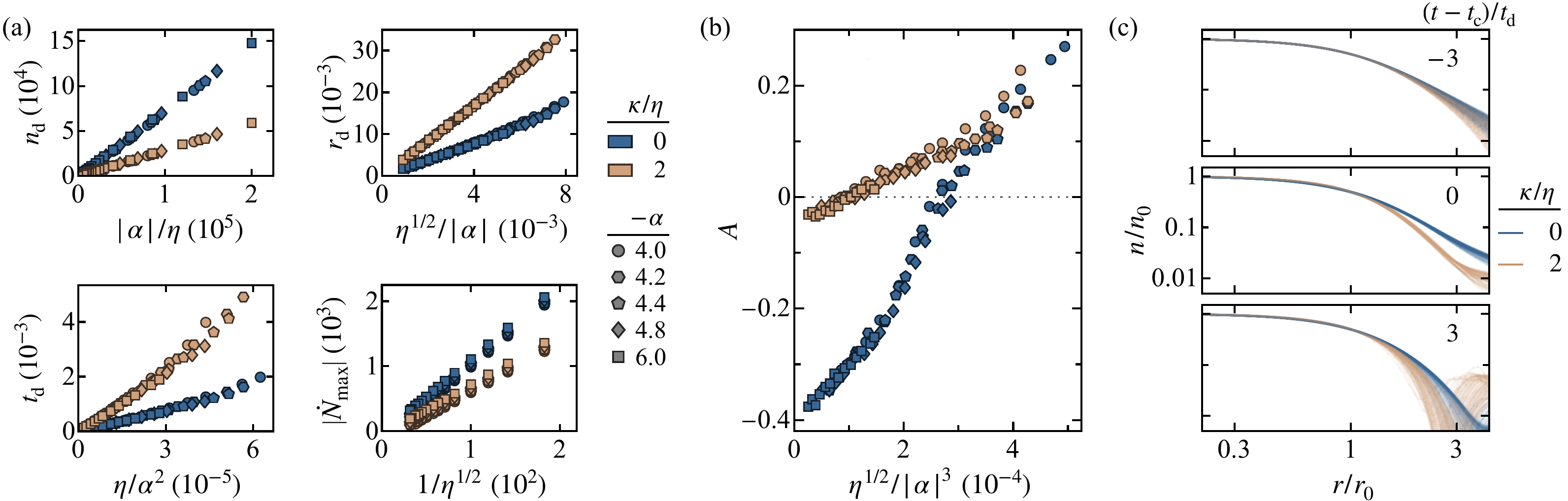}}
\caption{
Additional details on the effects of elastic three-body interactions.
(a)~The dissipation scales for both $\kappa=0$ and $\kappa/\eta=2$, also varying $\alpha$ and $\eta$ values as in Fig.~\ref{fig4}(a).
(b)~The loss asymmetry $A$ versus $\eta^{1/2}/|\alpha|^3$ for both $\kappa=0$ and $\kappa/\eta=2$, varying $\alpha$ and $\eta$ as in (a).
(c)~The normalized density profiles at three characteristic $(t-\tc)/\td$ (panels) for $\kappa=0$ and $\kappa/\eta=2$, varying $\alpha$ and $\eta$ values as in (a).
}
\label{figS4}
\end{figure*}

\newpage
\twocolumngrid


\begin{thebibliography}{81}
\bibitem [{tcv()}]{tcvsts}%
  \BibitemOpen
  \href@noop {} {}\bibinfo {note} {In the Zakharov weak-collapse
  picture, when the collapsing part of the wave is lost at a time $\tilde{t}$,
  $n_0 \propto |\alpha|/\eta \propto 1/[|\alpha|(\tinf - \tilde{t})]$, such
  that $(\tinf - \tilde{t}) \propto \eta/\alpha^2 > 0$. In our simulations, we
  find that $n_0$ first peaks at a time $\tc > \tinf$, because some dissipation occurs already before $\tilde{t}$ and slows down the growth of $n_0$.}\BibitemShut {Stop}%
\bibitem [{alp()}]{alphackappashift}%
  \BibitemOpen
  \href@noop {} {}\bibinfo {note} {{Note that for $\alpha=-4.0$, $\kappa/\eta=2$, and our two largest $\eta$ values, collapse does not occur,
  because increasing $\kappa$ also reduces $\ac$~\cite{Abdullaev:2001}; we
  numerically obtain $\ac \approx -3.89 - 40 \kappa$.}}\BibitemShut {Stop}%
\end{thebibliography}
\end{document}